\documentclass[12pt]{article}
\usepackage{amsfonts}
\usepackage{amssymb}
\def \ub {\underline}

\begin{document}

\setcounter{page}{0}

\renewcommand{\theequation}{{\thesection{.}}\arabic{equation}}

\rightline{\bf{ETH-TH/96-53\ }}

\vspace{1.5cm}

\centerline{\large{\bf{On Zero--Mass Ground States in Super--}}}

\centerline{\large{\bf{Membrane Matrix Models}}}

\vspace{2.5cm}

\large

\centerline{J\"urg Fr\"ohlich \ and \ Jens Hoppe\footnote{Heisenberg
    Fellow\\
\phantom{\quad \ } On leave of absence from Karlsruhe University}}

\centerline{Theoretical Physics}

\centerline{ETH-H\"onggerberg}

\centerline{CH--8093 \ Z\"urich}

\vspace{8cm}

\normalsize

\baselineskip=.65cm

\noindent {\bf Abstract} \ \  We recall a formulation of super-membrane
theory in terms of certain matrix models. These models are known to
have a mass spectrum given by the positive half-axis. We show that,
for the simplest such matrix model, a normalizable zero-mass ground
state does \underbar{not} exist.

\vfill\eject

\setcounter{page}{1}
\setcounter{equation}{0}%

\section{Introduction and Summary of Results}

Some time ago [1], super-membranes in $D$ space-time dimensions were
related to supersymmetric matrix models where, in a Hamiltonian
light-cone formulation, the $D-2$ transverse space coordinates appear
as non-commuting matrices [2].

It has been proven in [3] that the mass spectrum of any one of these
matrix models, which is given by the (energy) spectrum of some
supersymmetric 
quantum-mechanical Hamilton operator [4], fills the positive
half-axis of the real line. This property of the mass spectrum in
super-membrane models is in contrast to the properties of mass spectra
in bosonic membrane  matrix models [2] which are purely discrete; see
[5]. One of the important open questions concerning super-membrane
matrix models is whether they have a normalizable zero-mass ground
state. Such states would describe multiplets of zero-mass one-particle
states, including the gravitation; (see [1]). A new interpretation of
the mass spectrum of super-membrane matrix models (in terms of
multi-membrane configurations) has been proposed in [6].

A first step towards answering the question of whether there are
normalizable zero-mass ground states in super-membrane matrix models
has been undertaken in [1]. 
In this note, we continue the line of
thought described in [7] and show that, in the simplest matrix model,
a normalizable zero-mass ground state does not exist.

Let us recall the definition of super-membrane matrix models. The
configuration space of the bosonic degrees of freedom in such models
consists of $D-2$ copies of the Lie-algebra of $SU(N)$, for some
$N<\infty$, where $D$ is the dimension of space-time, with
$D=4,5,7,11$. A point in this configuration space is denoted by
$X=(X_j)$ with
\begin{equation}
X_j\;=\;\sum_{A=1}^{N^2-1} \ X_j^A \ T_A, \ j=1,\cdots, D-2, 
\end{equation}
where $\{ T_A\}$ is a basis of $su(N)$, the Lie algebra of $SU(N)$. In
order to describe the quantum-mechanical dynamics of these degrees of
freedom, we make use of the Heisenberg algebra generated by the
configuration space coordinates $X_j^A$ and the canonically conjugate
momenta $P_j^A$ satisfying canonical commutation relations
\begin{eqnarray}
\left[ X_j^A, \ X_k^B\right] &=& \left[ P_j^A, \ P_k^B\right] \ = \ 0,
\\ \nonumber
\left[ X_j^A, \ P_k^B\right] &=& i\;\delta^{\;AB} \ \delta_{jk} \ .
\end{eqnarray}

To describe the quantum dynamics of the fermionic degrees of freedom,
we make use of the Clifford algebra with generators $\Theta_\alpha^A$,
\ $A=1,\cdots,N^2-1,$ \ $\alpha=1,\cdots,2^{\ \left[
    \frac{D-2}{2}\right]}$, \  and commutation relations 
\begin{equation}
\left\{ \Theta_\alpha^A, \ \Theta_\beta^B\right\} \ = \
\delta_{\alpha\beta} \ \delta^{AB} \ .
\end{equation} 
The generators $\Theta_\alpha^A$ can be expressed in terms of
fermionic creation- and annihilation operators:
\begin{eqnarray}
\Theta_{2\alpha-1}^A &=& \frac{b_\alpha^A \ + \ c_\alpha^A}{\sqrt{2}}
\ , \nonumber \\ 
\Theta_{2\alpha}^A &=& \frac{i\, \left( b_\alpha^A -
    c_\alpha^A\right)}{\sqrt{2}}  \ ,
\end{eqnarray}
with \ $\left( b_\alpha^A\right)^* = c_\alpha^A$, \
$\alpha=1,\cdots,\frac 1 2 \ 2^{\left[ \frac{D-2}{2}\right]}$, \ and
\begin{eqnarray}
\left\{ b_\alpha^A, \ b_\beta^B\right\} &=& \left\{ c_\alpha^A, \
  c_\beta^B\right\} \ = \ 0 , \nonumber \\
\left\{ b_\alpha^A, \ c_\beta^B\right\} &=& \delta_{\alpha\beta} \
\delta^{AB}\ . 
\end{eqnarray}
The Hilbert space, ${\mathcal H}$, of state vectors (in the
Schr\"odinger representation) is a direct sum of subspaces ${\mathcal
  H}_k$, \ $k=0,\cdots,$ \ $K:=(N^2-1) \;\frac 1 2\;
  2^{\left[\frac{D-2}{2}\right]}$ . \ A vector $\Psi \in {\mathcal
  H}_k$ is given by
\begin{equation}
\Psi \ = \ \sum_{k=0}^K \ \frac{1}{k!} \ b_{\alpha_1}^{A_1} \cdots b_{\alpha_k}^{A_k} \
\psi_{A_1\cdots A_k}^{\alpha_1\cdots \alpha_k} \ (X) , 
\end{equation}
where $X=\left\{ X_j^A\right\}$, \ $j=1,\cdots, D-2$, $A=1,\cdots,
N^2-1$. We require that
\begin{equation}
c_\alpha^A \ \Psi\;=\;0, \quad {\rm for \ all} \quad \Psi \in {\mathcal
  H}_0\ .
\end{equation}
The scalar product of two vectors, $\Psi$ and $\Phi$, in ${\mathcal
  H}$ is given by 
\begin{equation}
\langle \Psi, \Phi\rangle\;=\;\sum_{k=0}^K\ \frac{1}{k!} \
\sum_{\alpha_1 \cdots \alpha_k \atop A_1 \cdots A_k} 
\int \prod_{j,A} dX_j^A \ \overline{\psi_{A_1\cdots
    A_k}^{\alpha_1\cdots \alpha_k} (X)} \ \times \ \phi_{A_1\cdots
  A_k}^{\alpha_1\cdots \alpha_k} (X) .
\end{equation}
The Hilbert space ${\mathcal H}$ carries unitary representations of
the groups $SU(N)$ and $SO(D-2)$. Let ${\mathcal H}^{(0)}$ denote the
subspace of ${\mathcal H}$ carrying the
trivial representation of $SU(N)$.

One can define supercharges, $Q_\alpha$ and $Q_\alpha^\dagger$, \
$\alpha=1,\cdots,\frac 1 2 \ 2^{\left[ \frac{D-1}{2}\right]}$, \ with
the properties that, on the subspace ${\mathcal H}^{(0)}$, 
\begin{equation}
\left\{ Q_\alpha, \ Q_\beta\right\}\biggm|_{{\mathcal H}^{(0)}} \ = \
\left\{ Q_\alpha^\dagger, \ Q_\beta^\dagger\right\}\biggm|_{{\mathcal H}^{(0)}} \ =
0 \ ,
\end{equation}
and
\begin{equation}
\left\{ Q_\alpha, \ Q_\beta^\dagger\right\}\biggm|_{{\mathcal H}^{(0)}} \ = \
\delta_{\alpha\beta} \ H\biggm|_{{\mathcal H}^{(0)}} \ ,
\end{equation}
where $H=M^2$, and $M$ is the mass operator of the super-membrane
matrix model.

Precise definitions of the supercharges and of the operator $H$ can be
found in [1] (formulas (4.7) through (4.12)). In [3] it is shown that
the spectrum of 
$H\bigm|_{{\mathcal H}^{(0)}}$ consists of the positive half-axis
$[0,\infty)$. The problem addressed in this note is to determine
whether $O$ is an eigenvalue of $H$ corresponding to a normalizable
eigenvector $\Psi_0 \in {\mathcal H}^{(0)}$. Using eqs.~(1.9) and
(1.10), one can show that $\Psi_0$ must be a solution of the equations
\begin{equation}
Q_\alpha \Psi \ = \ Q_\alpha^\dagger \Psi = 0,\quad {\rm for \ some} \
\alpha, \ \Psi \in {\mathcal H}^{(0)}\ .
\end{equation}
If eqs.~(1.11) have a solution, $\Psi_0=\Psi_{\alpha_0}$, for
$\alpha=\alpha_0$, they have a solution for all values of $\alpha$,
(by $SO(D-2)$ covariance). The problem to determine whether
eqs.~(1.11) have a solution, or not, can be understood as a problem
about the cohomology groups determined by the supercharges
$Q_\alpha$. We define
\[
{\mathcal H}_+ \ :=\ \displaystyle\mathop{\oplus}_{l\,\geq\,0} \
{\mathcal H}_{2 l}^{(0)}, \qquad {\mathcal H}_- \ :=\
\displaystyle\mathop{\oplus}_{l\,\geq\,0} \ {\mathcal H}_{2l+1}^{(0)} \ .
\]
We define the cohomology groups
\[
H_{\sigma,\alpha} \ :=\ \left\{ \Psi \in {\mathcal H}_\sigma^{(0)}
  \bigm| Q_\alpha \ \Psi=0\right\} \ \big/ \ \left\{ \Psi \mid \Psi =
  Q_\alpha \Phi, \Phi \in {\mathcal H}_{-\sigma}^{(0)} \right\} ,
\]
$\sigma = \pm 1$. If $H_{\sigma,\alpha}$ is non-trivial, for some
$\sigma$ and some $\alpha$, then eqs.~(1.11) have a solution.

\vspace{1cm}

\section{The $(D=4, N=2)$ model}

\setcounter{equation}{0}%

The goal of this note is a very modest one: We show that, for $D=4$
and $N=2$, eqs.~(1.11) 
do \underbar{not} have any normalizable solutions. Our proof is not conceptual;
it relies on explicit calculations and estimates and does therefore
not extend to the general case in any straightforward way.

When $D=4$ and $N=2$ we use the following notations:
\begin{eqnarray}
\vec{q}_j &:=& \left( X_j^1, X_j^2, X_j^3\right), \quad
j=1,2,\nonumber\\
\vec{\lambda} &=&\left( \lambda^1, \lambda^2, \lambda^3\right)\ :=\
\left( b_\alpha^1, \ b_\alpha^2, \ b_\alpha^3\right),
\end{eqnarray}
and
\[
\frac{\partial}{\partial\vec{\lambda}}\ =\
\left(\frac{\partial}{\partial\lambda^1},\
  \frac{\partial}{\partial\lambda^2},\
  \frac{\partial}{\partial\lambda^3}\right) \ :=\ \left( c_\alpha^1,
  c_\alpha^2, c_\alpha^3\right),
\]
$\alpha=1$. \ The operators representing the generators of $su(2)$ on
${\mathcal H}$ are given by
\begin{equation}
\vec{L}\ = \ - i\;  \left( \vec{q}_1 \wedge \vec{\nabla}_1 + \vec{q}_2 \wedge
  \vec{\nabla}_2 + \vec{\lambda} \wedge
  \frac{\partial}{\partial\vec{\lambda}} \right) \ .
\end{equation}
The supercharges are given by (see [1], eq.~(4.20))
\begin{eqnarray}
Q &=& \left( \vec{\nabla}_1 - i\,\vec{\nabla}_2\right) \;\cdot\;
\frac{\partial}{\partial\vec{\lambda}} \ + \ i\;\vec{q}\;\cdot\;
\vec{\lambda}\ ,\nonumber \\ 
{\rm and} \quad &\quad & \\
Q^\dagger &=& -\;\left(\vec{\nabla}_1 + i\, \vec{\nabla}_2\right)
\;\cdot\;\vec{\lambda} \,-\, i\, \vec{q} \; \cdot\;
  \frac{\partial}{\partial\vec{\lambda}} \ , \nonumber
\end{eqnarray}
where
\begin{equation}
\vec{q} \ = \ \vec{q}_1 \wedge \vec{q}_2 \ ,
\end{equation}
and $\wedge$ denotes the vector product. We then have that
\[
Q^2\ = \ \left( \vec{q}_1-i\,\vec{q}_2\right)\;\cdot\; \vec{L}\; , \
(Q^\dagger)^2\ =\ \left( \vec{q}_1 + i\,\vec{q}_2\right) \;\cdot\;\vec{L} ,
\]
and
\begin{equation}
H \ = \ \left\{ Q,\; Q^\dagger\right\} \ .
\end{equation}
A vector $\Psi \in {\mathcal H}_+$ can be written as
\begin{eqnarray}
\Psi &=& \psi + \frac 1 2 \ \left( \vec{\lambda} \wedge
  \vec{\lambda}\right) \; \cdot \;\vec{\psi} \nonumber\\
 &=& \psi + \frac 1 2 \ \varepsilon_{ABC} \
 \lambda^A\;\lambda^B\;\psi^C \ .
\end{eqnarray}
For $\Psi\in{\mathcal H}_+^{(0)}$ \ (i.e., $\Psi\in{\mathcal H}_+$
with $\vec{L}\Psi = 0$), eqs.~(1.11) imply the following system $(^*)$
of first-order differential equations:
\begin{eqnarray}
i\;\vec{q}\;\psi &=& \left( \vec{\nabla}_1 -
  i\,\vec{\nabla}_2\right) \wedge \vec{\psi}\ , \\
\vec{q}\;\cdot\;\vec{\psi} &=& 0 \ ,
\end{eqnarray}
and
\begin{eqnarray}
\left( \vec{\nabla}_1 + i\,\vec{\nabla}_2\right) \ \psi &=& i\;\vec{q}
\wedge\vec{\psi} \ , \\
\left( \vec{\nabla}_1 + i\,\vec{\nabla}_2\right)\;\cdot\;\vec{\psi}
&=& 0 \ .
\end{eqnarray}
Moreover, the equation \ $\vec{L} \Psi = 0$ yields
\begin{eqnarray}
&&\left(\vec{q}_1\wedge\vec{\nabla}_1+\vec{q}_2\wedge\vec{\nabla}_2\right)
\; \psi \ = \ 0\ , \\
&& \left( \vec{q}_1 \wedge \vec{\nabla}_1 + \vec{q}_2 \wedge
  \vec{\nabla}_2\right)_A\; \psi_B\ + 
\sum_C \ \varepsilon_{ABC}\;\psi_c\ = \ 0 \ , \quad \forall\; A,B \ .
\end{eqnarray}
It is straightforward to verify that, for \ $\Psi\in{\mathcal
  H}_-^{(0)}$, \ eqs.~(1.11) imply a system of equation equivalent to
(2.7) through (2.12). This can be interpreted as a consequence of
Poincar\'e duality.

The formal expression for the Hamiltonian \ $H = \left\{ Q,
  Q^\dagger\right\}$ \ is given by
\begin{equation}
H \ = \ H_B \ + \ H_F \ ,
\end{equation}
where
\[
H_B\ = \ -\;\vec{\nabla}_1^2\;-\;\vec{\nabla}_2^2\;+\;\vec{q}^{\,2}
\]
and
\begin{equation}
H_F\ = \ \left( \vec{q}_1 + i\,\vec{q}_2\right)\;\cdot\; \left(
  \vec{\lambda} \wedge \vec{\lambda}\right) - \left( \vec{q}_1 -
  i\,\vec{q}_2\right) \cdot
\left(\frac{\partial}{\partial\vec{\lambda}} \wedge
  \frac{\partial}{\partial\vec{\lambda}}\right) \ .
\end{equation}
As shown in [5], the spectrum of $H_B$ is discrete, with
\begin{equation}
{\rm inf \ spec} \  H_B \ = \ E_0 \;>\; 0 \ .
\end{equation}
The representation of the group $SO(D-2)\simeq U(1), (D=4)$ on
${\mathcal H}$ is generated by the operator
\begin{equation}
J\ = \ -\,i\;\left( \vec{q}_1
  \;\cdot\;\vec{\nabla}_2\;-\;\vec{q}_2\;\cdot\;\vec{\nabla}_1\right)
\ - \ \frac 1 2 \
\vec{\lambda}\;\cdot\;\frac{\partial}{\partial\vec{\lambda}} \ .
\end{equation}
While $J$ does not commute with $Q$ or $Q^\dagger$, it does commute with
$QQ^\dagger$ and $Q^\dagger Q$ and hence with $H$. It is therefore
sufficient to 
look for solutions of eqs.~(2.7) through (2.12) transforming under an
irreducible representation of $U(1)$, i.e. solutions that are
eigenvectors of $J$ corresponding to eigenvalues \ $j \in \frac 1 2 \
{\mathbb Z}$ . \ The spectrum of the restriction of $J$ to the
subspace ${\mathcal H}_+$ is the integers, while \ spec
$\left( J\bigm|_{{\mathcal H}_-}\right)$ consists of half-integers.

\vspace{1cm}

\section{Analysis of equations \ $(^*)$}
\setcounter{equation}{0}%

In this section, we assume that $Q\Psi= Q^\dagger\Psi=0$ has a solution
$\Psi\in{\mathcal H}_+^{(0)}$ and then show that $\Psi=0$. 

The assumption that $Q\Psi = Q^\dagger\Psi=0$ implies that
\begin{equation}
\langle Q\Psi, Q\Psi\rangle + \langle Q^\dagger\Psi, Q^\dagger\Psi\rangle\ = \ 0\ .
\end{equation}
Let $\xi := \left(\vec{q}_1,\vec{q}_2\right) \in{\mathbb R}^6$.
Let $g_n(\xi)\equiv g_n(|\xi |)$, $n=1,2,3,\cdots,$ \ be a function on
${\mathbb R}^6$ only depending on \ $|\xi| := \sqrt{\vec{q}_1^{\,2} +
  \vec{q}_2^{\,2\,}}$ \ with the properties that $g_n$ is smooth,
monotonic decreasing, 
$g_n(|\xi|)=1$, for $|\xi| \leq n$, $g_n(|\xi|)=0$ \ for $|\xi|\geq
3n$, and \ $\left|\left(\frac{d}{dt}\; g_n\right) (t)\right| \ \leq \
\frac 1 n$ . Let $h_k (\xi)$, $k=1,2,3,\cdots,$ be an approximate
$\delta$-function at $\xi=0$ with the properties that $h_k$ is smooth,
$h_k \geq 0$, $\int h_k (\xi) d^6 \xi = 1$, and
\begin{equation}
{\rm supp} \ h_k \ \subseteq \ \left\{ \xi \biggm| \ |\xi| \ \leq \
  \frac{1}{k^2} \right\} \ . 
\end{equation}
We define a bounded operator, $R_{n,k}$, on ${\mathcal H}$ by setting
\begin{equation}
\left( R_{n,k} \Phi\right)\left(\xi\right)\ = \ g_n \left(\xi\right)
\int h_k \left( \xi -\xi'\right) \Phi \left(\xi'\right) d^6 \xi' \ ,
\end{equation}
for any $\Phi \in {\cal H}$. Clearly
\begin{equation}
\displaystyle\mathop{s-lim}_{n\to\infty \atop k\to\infty} \ R_{n,k}
\;\Phi\;=\;\Phi\ ,
\end{equation}
for any $\Phi\in{\cal H}$. Next, we note that, for a vector $\Phi$ in
the domain of the operator $Q$,
\begin{equation}
\left(\left[ Q,
    R_{n,k}\right]\Phi\right)\left(\xi \right)\;=\;I_{n,k}
\left(\xi\right)\;+\; I\!\!I_{n,k}\left(\xi\right),  
\end{equation}
where
\begin{equation}
I_{n,k}\left(\xi\right)\;=\;\left(\left(\vec{\nabla}_1-i\vec{\nabla}_2\right)
  g_n\right)\left(\xi\right)\cdot\int h_k\left(\xi-\xi'\right)\;
\frac{\partial}{\partial \vec{\lambda}}\;\Phi \left(\xi'\right)
d^6\xi'\ ,
\end{equation}
and
\begin{equation}
I\!\!I_{n,k}\left(\xi\right)\;=\;i\,g_n\left(\xi\right) \int h_k
\left(\xi-\xi'\right) \left(\vec{q}\left(\xi\right)-\vec{q}
  \left(\xi'\right)\right) \cdot\vec{\lambda} \Phi \left(\xi'\right)
d^6 \xi' \ .
\end{equation}
The operator norm of the operators \
$\frac{\partial}{\partial\lambda^A}$ \ and $\lambda^A$, $A=1,2,3,$ is
bounded by 1. Since \ $\left| \frac{d}{dt}\;g_n (t)\right| \leq \frac
1 n$, \ the operator norm of the multiplication operator \
$\left(\left(\vec{\nabla}_1 - i \vec{\nabla}_2\right)
  g_n\right)\left(\cdot\right)$ \ is bounded above by \ $\frac 1 n $ \
. The operator norm of the convolution operator \ $\Phi\left(\xi\right)
\mapsto \int h_n \left(\xi-\xi'\right) \Phi \left(\xi'\right)
d^6\xi'$ \ is equal to 1.
This implies
\begin{equation}
\Vert I_{n,k}\Vert \;\leq\;\frac 6 n \;
\bigg\Vert \frac{\partial}{\partial\vec{\lambda}}\;\Phi\bigg\Vert
\;\leq\;\frac{18}{n}\; \Vert \Phi \Vert \ .
\end{equation}
Next, we note that, for $\xi$ in the support of the function $g_n$,
$$
\biggl| h_k\left( \xi -\xi'\right) 
           \left( \vec{q}\left( \xi \right) -\vec{q}\left(
               \xi'\right)\right) \biggr| \;\leq\; \frac{7 n}{k^2} \;
          h_k \left( \xi -\xi'\right) \ .
$$
Thus, for $k\geq n$
\begin{equation}
\Vert I\!\!I_{n,k}\Vert \;\leq\;\frac{21}{n}\; \Vert\Phi\Vert \ .
\end{equation}
In conclusion
\begin{equation}
\Vert \left[ Q, R_{n,k} \right] \ \Phi \Vert \ \leq \ \frac{40}{n} \
\Vert\Phi\Vert \ ,
\end{equation}
for $k\geq n$.

\noindent
A similar chain of arguments shows that, for $\Phi$ in the domain of
$Q^\dagger$,
\begin{equation}
\Vert \left[ Q^\dagger, R_{n,k}\right] \ \Phi\Vert \ \leq \ \frac{40}{n}\
\Vert\Phi\Vert \ ,
\end{equation}
for $k\geq n$.

Next, we suppose that $\Psi$ solves (3.1). We claim that, given
$\varepsilon  >0$, there is some finite $n(\varepsilon)$ such that,
for $\Psi_{n,k} := R_{n,k}\Psi$ ,
\begin{equation}
\Vert \Psi\Vert\;\geq\;\Vert\Psi_{n,k}\Vert\;\geq\;\left(
  1-\varepsilon\right) \Vert\Psi\Vert \ ,
\end{equation}
and
\begin{equation}
\langle Q\Psi_{n,k}, Q\Psi_{n,k}\rangle\;+\; \langle Q^\dagger\Psi_{n,k},
Q^\dagger\Psi_{n,k}\rangle \;\leq\;\varepsilon\Vert\Psi\Vert^2 \ ,
\end{equation}
for all $k\geq n\geq n(\varepsilon)$. \ Inequality (3.12) follows
directly from (3.4) and the fact that the operator norm of \ $R_{n,k}$
is $=1$. To prove (3.13), we note that, for $k\geq n$,
\begin{equation}
\langle Q^\# \Psi_{n,k}, Q^\# \Psi_{n,k}\rangle \;=\;
\langle \left[ Q^\#,R_{n,k}\right]\Psi, \left[ Q^\#, R_{n,k}\right]
\Psi\rangle \;\leq\;
\left(\frac{40}{n}\right)^2 \ \langle \Psi,\Psi\rangle \ ,
\end{equation}
where \ $Q^\#=Q$ or $Q^\dagger$. This follows from the equations $Q\Psi =
Q^\dagger\Psi = 0$ and inequalities (3.10) and (3.11).

We now observe that, by the definition of $R_{n,k}$,
$\Psi_{n,k}=R_{n,k}\Psi$ is a smooth function of compact support in
${\mathbb R}^6$, for all $n\leq k <\infty$. It 
therefore belongs to the domain of definition of the operators \
$Q\;Q^\dagger$ and $Q^\dagger\;Q$. Thus, for all $n \leq k <\infty$, 
\begin{eqnarray}
&& \langle Q \Psi_{n,k}, Q\Psi_{n,k}\rangle \;+\;
\langle Q^\dagger \Psi_{n,k}, Q^\dagger\Psi_{n,k}\rangle \nonumber \\
&& \qquad = \; \langle \Psi_{n,k}, \left\{ Q, Q^\dagger\right\} \
\Psi_{n,k}\rangle \\
&& \qquad = \; \langle \Psi_{n,k}, H_B \Psi_{n,k}\rangle \;+\;
\langle \Psi_{n,k}, H_F \Psi_{n,k}\rangle \ ,\nonumber
\end{eqnarray}
where $H_B$ and $H_F$ are given in eq.~(2.14), (and it is obvious from
(2.14) that $\Psi_{n,k}$ belongs to the domains of definition of $H_B$
and $H_F$).

As proven in [5],
\begin{equation}
\langle \Phi, H_B \Phi\rangle \;\geq \; E_0 \Vert\Phi\Vert^2 \ ,
\end{equation} 
for some strictly positive constant $E_0$ (= inf spec $H_B$), for all
vectors $\Phi$ in the domain of $H_B$. Thus, for $k\geq n \geq n
(\varepsilon)$, and using (3.13), we have that
\begin{eqnarray}
\varepsilon\;\Vert\Psi\Vert^2 &\geq& 
\langle\Psi_{n,k}, H_B \Psi_{n,k}\rangle \ + \ 
\langle\Psi_{n,k}, H_F \Psi_{n,k}\rangle \nonumber\\
&\geq& \left( 1-\varepsilon\right)^2 E_0 \Vert\Psi\Vert^2\;+\;
\langle\Psi_{n,k}, H_F \Psi_{n,k}\rangle \ .
\end{eqnarray}
Our next task is to analyze \ $\langle\Psi_{n,k}, H_F
\Psi_{n,k}\rangle$. If $\Phi = \left( \varphi, \vec{\varphi}\right)
\in {\mathcal H}_+$ \ belongs to the domain of definition of $H_F$
then
\begin{equation}
\langle \Phi, H_F\Phi\rangle \;=\; 2 \int \
\overline{\varphi\left(\xi\right)\,} \ \left(\vec{q}_1 -
  i\,\vec{q}_2\right) \cdot \vec{\varphi} \left(\xi\right) \ d^6\xi
\;+\; c.c.
\end{equation}
Note that $\Psi_n := \displaystyle\mathop{\rm lim}_{k\to\infty} \
\Psi_{n,k}$, \ where $\Psi_{n,k}=R_{n,k}\Psi$ and $\Psi$ solves (3.1),
belongs to the domain of definition of $H_F$. Since $\Psi = \left(
  \psi,\vec{\psi}\right)$ solves the equations $Q\Psi = Q^\dagger\Psi=0$,
see (3.1), we can use eqs.~(2.8) and (2.9) to eliminate $\vec{\psi}$:
For $\vec{q} \neq 0$, we find that
\begin{equation}
\vec{\psi}\left(\xi\right)\;=\;\frac{i\,\vec{q}}{q^2} \ \wedge \
\left( \vec{\nabla}_1 \;+\;i\,\vec{\nabla}_2\right) \ \psi (\xi) 
\end{equation}
(recall that $\vec{q}=\vec{q}_1
\wedge\vec{q}_2$). Inserting (3.19) on the R.S. of (3.18), for
$\Phi=\Psi_n$, we arrive at the equation
\begin{eqnarray}
\langle\Psi_n, H_F \Psi_n\rangle &=&
2\;\int\overline{\left(g_n\psi\right)\left(\xi\right)} \left(\vec{q}_1
  - i\,\vec{q}_2\right) \nonumber\\
&& \qquad \times \ g_n \left(\xi\right) \left( \frac{i\,\vec{q}}{q^2}
  \;\wedge\;\left(
    \vec{\nabla}_1+i\,\vec{\nabla}_2\right)\;\psi\right)
\left(\xi\right) \; d^6\xi\;+\;c.c.
\end{eqnarray}
Let 
$$
T \ := \ 2 \left( \vec{q}_1 \;-\;i\,\vec{q}_2\right) \cdot \left(
  \frac{i\,\vec{q}}{q^2}\;\wedge\;\left(\vec{\nabla}_1\;+\;
    i\,\vec{\nabla}_2\right)\right) \ .  
$$
Then
\begin{eqnarray}
\langle\Psi_n, H_F \Psi_n\rangle &=& \langle \Psi_n,
\;T\;\Psi_n\rangle\;+\; c.c. \nonumber \\
&-& \int \mid\overline{\psi\left(\xi\right)}\mid^2 \;
g_n\left(\xi\right) \left[ T, g_n\right]\left(\xi\right)\ d^6\xi\;+\;c.c.
\end{eqnarray}
Next, we make use of the fact that $\Psi$ must be
SU(2)--invariant. This is expressed in eq.~(2.11), which 
implies that $\psi(\xi)$ only depends on SU(2)--invariant combinations
of the variables $\vec{q}_1$ and $\vec{q}_2$, i.e., on
\begin{equation}
r_1\;:=\;|\vec{q}_1|, \ r_2\;:=\;|\vec{q}_2|, \ x\;:=\;
\frac{\vec{q}_1\cdot\vec{q}_2}{r_1\;r_2} \ .
\end{equation}
Instead, we may use variables $q,p$ and $\varphi$ defined by
\begin{equation}
p\,e^{i\varphi}\;:=\;\frac 1 2 \
\left(\vec{q}_1\;+\;i\,\vec{q}_2\right)^2\;=\;\frac 1 2 \ \left(
  r_1^2\;-\;r_2^2\right)\;+\; i\,r_1r_2 \;x, \quad  q\;:=\;
|\vec{q}_1\wedge\vec{q}_2| 
\end{equation}
with 
\begin{equation}
0\;\leq\;p\;<\;\infty, \quad
0\;\leq\;\varphi\;<\;4\pi,  \quad
0\;\leq\;q\;<\;\infty \ . 
\end{equation}
If $F$ is an SU(2)--invariant function then
\begin{equation}
\int d^6\xi\;F\;\left(\xi\right)\;=\;c \int\limits_0^\infty dq
\int\limits_0^\infty dp \int\limits_0^{4\pi} d\varphi \
\frac{q\,p}{\sqrt{q^2\,+\,p^2\,}} \;F\; \left( q,p,\varphi\right) \ ,
\end{equation}
where $c$ is some positive constant.

If $\varphi$ is SU(2)--invariant then 
\begin{equation}
T\,\varphi\;=\;c' \ \frac{\sqrt{q^2 + p^2}}{q} \
\frac{\partial}{\partial\,q} \ \varphi \ ,
\end{equation}
where $c'$ is a positive constant. 

Using (3.26) in (3.21), we find that
\begin{eqnarray}
&&\langle\Psi_n,H_F\Psi_n\rangle\;=\;c'' \int\limits_0^\infty dp
\int\limits_0^{4\pi} d\varphi \int\limits_0^\infty dq\;p\
\frac{\partial}{\partial q} \ |\psi_n \left(
  q,p,\varphi\right)|^2\nonumber \\
&&\qquad -\;c'' \int\limits_0^\infty dp \int\limits_0^{4\pi} d\varphi
\int\limits_0^\infty dq \; p \ |\psi \left( q,p,\varphi\right)|^2 \
\frac{\partial}{\partial q} \ \left( g_n \left(
    2\;\sqrt{q^2+p^2}\right)^2\right)\ ,
\end{eqnarray}
with $c'' = c.c' \;>\;0$.

By the definition of $g_n$,
$$
\frac{\partial}{\partial q} \ \left( g_n \left(
    2\;\sqrt{q^2+p^2}\right)^2\right) \;\leq\;0 \ ,
$$
pointwise. Therefore
\begin{equation}
\langle \Psi_n, H_F \Psi_n\rangle\;\geq\;-\;c'' \int\limits_0^\infty
dp \int\limits_0^{4\pi} d\varphi\;p\; |\psi_n \left( q=0, p,
  \varphi\right)|^2 \ .
\end{equation}
In passing from (3.27) to (3.28), we have used that \
$\frac{\partial}{\partial q} |\psi_n \left( q,p,\varphi\right)|^2$ \
is an $L^1$--function with respect to the measure \
$p\;dp\;d\varphi\;dq$ \ and that
$$
\int\limits_0^\infty dp \int\limits_0^{4\pi} d\varphi\;p\;|\psi_n
\left( q,p,\varphi\right)|^2 
$$
is right-continuous at $q=0$. These facts will be proven below.

Combining eqs.~(3.15), (3.17) and (3.28), we conclude that 
\begin{eqnarray}
\varepsilon \Vert\Psi\Vert^2 &\geq& \displaystyle\mathop{\rm
  lim}_{k\to\infty} 
\left\{ \langle Q \Psi_{n,k}, Q \Psi_{n,k}\rangle\;+\; \langle Q^\dagger
  \Psi_{n,k}, Q^\dagger\Psi_{n,k}\rangle\right\}\nonumber\\
&\geq& \left( 1-\varepsilon\right) E_0 \Vert\Psi\Vert^2 \\
&-& c'' \int\limits_0^\infty dp \int\limits_0^{4\pi} d\varphi
\;p\;|\psi_n \left( q=0, p,\varphi\right)|^2\ , \nonumber
\end{eqnarray}
for all $n\,\geq\,n(\varepsilon)$. \ Choosing $\varepsilon$
sufficiently small, we conclude that either $\Psi=0$, or there is a
constant $\beta >0$ such that
\begin{equation}
\int\limits_0^\infty dp \int\limits_0^{4\pi}
d\varphi\;p\;|\psi_n\left( q=0,p,\varphi\right) |^2 \ \geq \ \beta \ ,
\end{equation}
for all sufficiently large $n$.

Next, we explore the consequences of (3.30). Since $\Psi$ solves
(3.1), we can use (3.19) to conclude that 
\begin{eqnarray}
\infty\;>\;\Vert\Psi\Vert^2 &=&
\Vert\psi\Vert^2\;+\;\Vert\vec{\psi}\Vert^2\nonumber \\
&=& \int d^6\xi\;\left\{
  |\psi(\xi)|^2\;+\;\frac{|\left(\vec{\nabla}_1 + i\,
      \vec{\nabla}_2\right)\;\psi(\xi)|^2}{|\vec{q}|^2}\ \right\} \
\nonumber .  
\end{eqnarray}
Using that $\Psi$ is SU(2)--invariant and passing to the variables
$q,p$ and $\varphi$, one finds that
\begin{equation}
2\;\int\limits_0^\infty dp\;p \int\limits_0^\infty \frac{dq}{q}
\int\limits_0^{4\pi} d\varphi
\left(\biggm|\psi,_p\;+\;\frac{i\psi,_\varphi}{p}\;\biggm|^2\;+\;
\bigm|\psi,_q\bigm|^2 \right)\left( q,p,\varphi\right)\;<\;\widetilde{K} ,
\end{equation}
where $\psi,_x := \frac{\partial\psi}{\partial x}$ , and
\begin{equation}
\int\limits_0^\infty dp \int\limits_0^\infty dq \int\limits_0^{4\pi}
d\varphi\; \frac{pq}{\sqrt{p^2+q^2}} \left|
  \psi\left(q,p,\varphi\right)\right|^2 \;<\; \widetilde{K},
\end{equation}
with $\widetilde{K} = \frac{\Vert\Psi\Vert^2}{c} < \infty$ (with the
constant $c$ 
appearing in (3.25)). 

Inequalities (3.31) and (3.32) also hold for $\psi_n$, instead of
$\psi$, with a constant $K$ that is uniform in $n\to\infty$. These
inequalities prove that \ $\frac{\partial}{\partial
  q}\left|\psi_n\left(q,p,\varphi\right)\right|^2$ \ is an
$L^1$--function with respect to the measure \ $p\;dp\;d\varphi\;dq$ \
and that
$$
f_n (q)\;:=\;\int\limits_0^\infty dp \int\limits_0^{4\pi} d\varphi\,p
\left| \psi_n \left( q,p,\varphi\right)\right|^2
$$
is right-continuous at $q=0$, properties that were used in our
derivation of (3.28).

By the Schwarz inequality and (3.31),
\begin{eqnarray}
&& \int\limits_0^\infty dp \int\limits_0^{4\pi} d\varphi
\int\limits_0^\infty dq\; p \left| \frac{\partial}{\partial q}\left|
    \psi_n \left( q,p,\varphi\right)\right|^2\right| \nonumber\\
&& \leq\; 2 \left( \int\limits_0^\infty dp\,p \int\limits_0^\infty
  \frac{dq}{q} \int\limits_0^{4\pi} d\varphi \left| \psi_{n,q}
    \left(q,p,\varphi\right)\right|^2\right)^{1/2} \nonumber \\
&& \qquad \left( \int\limits_0^\infty dp\,p \int\limits_0^\infty q\,dq
  \int\limits_0^{4\pi} d\varphi \left| \psi_n\left(
      q,p,\varphi\right)\right|^2\right)^{1/2} \;\leq\; K' n^4 ,\nonumber
\end{eqnarray}
for some finite constant $K'$ !

To prove continuity of $f_n(q)$ in $q$, we note that, for $q_1> q_2$,
$$
\left| f_n\left( q_1\right)\;-\;f_n\left(q_2\right)\right|
\;\leq\; \int\limits_0^\infty dp \int\limits_0^{4\pi} d\varphi
\int\limits_{q_2}^{q_1} dq\;p\;\left| \frac{\partial}{\partial
    q}\left| \psi_n \left( q,p,\varphi\right) \right|^2 \right|
$$
with tends to 0, as $\left( q_1-q_2\right) \to 0$, because \
$\frac{\partial}{\partial q}\left| \psi_n \left(
    q,p,\varphi\right)\right|^2$ \ is an $L^1$--function.

Next, we make use of the $SO(D-2) \simeq U(1)$ symmetry with generator
$J$ given in eq.~(2.16). We have noted below (2.16) that $J$ commutes
with $Q Q^\dagger$ and $Q^\dagger Q$, and hence that $\Psi \in {\mathcal H}_+$ can
be chosen to be an eigenvector of $J$ corresponding to some eigenvalue
$m \in {\mathbb Z}$. In the variables $q,p,\varphi$, 
$$
J\;=\;-\,2\,i\; \frac{\partial}{\partial\varphi} \ .
$$
Hence we may write
\begin{equation}
\psi \left(q,p,\varphi\right) \;=\; e^{i\,\frac m 2 \, \varphi} \
p^{\frac m 2} \phi \left( q,p\right) , 
\end{equation}
for some function $\phi$ independent of $\varphi$. Eqs.~(3.31) and
(3.32) then simply
\begin{equation}
\int\limits_0^\infty dp\,p^\alpha \int\limits_0^\infty \frac{dq}{q}
\left(\left| \phi,_p \right|^2 + \left| \phi,_q\right|^2\right) \left(
  q,p\right) < \infty
\end{equation}
and
\begin{equation}
\int\limits_0^\infty dp \int\limits_0^\infty dq \; \frac{p^\alpha
  q}{\sqrt{p^2+q^2}}\; \left| \phi \left( q,p\right)\right|^2 < \infty ,
\end{equation}
where $\alpha = m +1$. Furthermore, inequality (3.30), in the limit as
$n\to\infty$, yields 
\begin{equation}
\int\limits_0^\infty dp\,p^\alpha \ \left| \phi \left(
    q=0,p\right)\right|^2 \;\geq \; \frac{\beta}{4\pi} \ .
\end{equation}
Let $I_N := \left[ \frac 1 N , N\right]$. Then inequality  (3.34)
implies that there exists a set \ $\Omega \subseteq \left[ 0, \infty
\right)$ with the property that 
$\Omega \cap \left[ 0,\delta\right]$ has Lebesgue measures $\frac
\delta 2$, for any $\delta > 0$, and such that
\begin{equation}
\left( \frac 1 N\right)^{|\alpha|} \int\limits_{I_N} dp \left| \phi,_p
  \left( q, p\right)\right|^2\;\leq\; K_\delta
\end{equation}
for some constant $K_\delta$ independent of $N$ and all
$q\in\Omega\cap \left[ 0,\delta\right]$. Moreover,
\begin{equation}
\displaystyle\mathop{\rm lim}_{q \to 0 \atop q \in \Omega}
\int\limits_{I_N} dp \left| \phi,_p \left( q,p\right)\right|^2 \;=\;0 ,
\end{equation}
for all $N$. It follows that, for $q\in \Omega \cap \left[
  0,\delta\right], \ p_1, p_2 \in I_N, \ N < \infty,$
\begin{equation}
\left| \phi\left( q,p_1\right) - \phi \left( q,p_2\right)\right|\;=\;
\left| p_1-p_2 \right| \left| \int\limits_{p_1}^{p_2} \frac{dp}{\left|
      p_1-p_2\right|} \phi,_p \left(q,p\right)\right| \;\leq\; \left|
  p_1-p_2\right|^{1/2} \left( N^{\left|\alpha\right|}
  K_\delta\right)^{1/2} .
\end{equation}
Thus, for $q\in\Omega\cap \left[ 0,\delta\right]$ and $p_1,p_2 \in
I_N$, $\phi \left( q,p\right)$ is uniformly H\"older--continuous with
exponent $\frac 1 2$. Thus
$\phi_0 (p) := \displaystyle\mathop{\rm lim}_{q\to 0 \atop q\in\Omega}
\phi (q,p)$ is uniformly continuous in $p\in I_N$, for all $N<\infty$,
and it then follows from (3.38) that
\begin{equation}
\phi_0 \left( p\right) \ = \ \phi_0 \ = \ {\rm const.}
\end{equation} 
Inequality (3.36) then implies that $|\phi_0|$ must be positive!
Without loss of generality, we may then assume that $\phi_0 > 0$.

Thus the function $\phi$ introduced in (3.33) has the following
properties

\begin{eqnarray}
&&(A)\quad \displaystyle\mathop{\rm lim}_{q\to 0 \atop q\in\Omega} \;
\phi\left( q,p\right) \;=\; \phi_0\;>\;0 \nonumber\\
&&(B)\quad \int\limits_0^\infty dp \int\limits_0^\infty dq \
\frac{qp^\alpha}{\sqrt{p^2+q^2}} \ \left| \phi\left(
    q,p\right)\right|^2 \;<\;\infty\nonumber\\
&&(C)\quad \int\limits_0^\infty dp \int\limits_0^\infty dq\
\frac{p^\alpha}{q} \ \left(\left| \phi,_p \left( q,p\right)\right|^2 +
  \left| \phi,_q \left( q,p\right)\right|^2\right) \;<\;\infty \ . \nonumber
\end{eqnarray}

\noindent We now show that such a function $\phi\left(q,p\right)$ does
\underbar{not} exist. 

Let us first consider the case $\alpha \geq 0$. We choose an
arbitrary, but fixed $p\in \left( 0,\infty\right)$. Using the Schwarz
  inequality, we find that, for $0<q_0<\infty$, 
\begin{eqnarray}
\int\limits_0^{q_0} \frac{dq}{q} \left| \phi,_q \left(
    q,p\right)\right|^2 
&\geq& \frac{1}{q_0} \int\limits_0^{q_0} dq \left| \phi,^q \left(
    q,p\right)\right|^2\nonumber \\
&\geq& \left( \frac{1}{q_0} \int\limits_0^{q_0} dq \left| \phi,_q
    \left( q,p\right)\right|\right)^2  \\
&\geq& \left| \frac{1}{q_0} \int\limits_0^{q^*(p)} \;dq\;\phi,_q
  \left( q,p\right)\right|^2 , \nonumber
\end{eqnarray}
where $q^* (p) \in [0,q_0]$ is the point at which $\left|\phi
  (q,p)\right|$ takes its minimum in the interval $[0,q_0]$. Note that
$\phi(q,p)$ is continuous in $q \in [0,q_0]$, for almost every $p\in
[0,\infty)$. The R.S. of (3.41) is equal to 
$$
\frac{1}{q_0^2} \; \left| \phi \left( q^* (p), p\right) -
  \phi_0\right|^2 \ .
$$
Thus
\begin{equation}
\left( \frac{1}{q_0} \left| \chi \left(p\right) -
    \phi_0\right|\right)^2 \ \leq \ \int\limits_0^{q_0} \frac{dq}{q}
\left| \phi,_q \left( q,p\right)\right|^2,
\end{equation}
where  $\chi (p) = \phi (q^* (p), p)$. By property (C),
$$
\int\limits_0^\infty dp\;p^\alpha \int\limits_0^{q_0} \frac{dq}{q}
\left| \phi,_q \left( q,p\right)\right|^2 \ \leq \ \varepsilon (q_0) ,
$$
for some finite $\varepsilon (q_0)$, will $\varepsilon (q_0) \to 0$,
as $q_0\to 0$. Hence
\begin{equation}
\int\limits_0^\infty dp\;p^\alpha \left| \chi (p) - \phi_0\right|^2 \
< \ q_0^2 \varepsilon \left( q_0\right). 
\end{equation}
We define a subset $M_\delta \subseteq [ 0,\infty )$ by
$$
M_\delta \ := \ \left\{ p \biggm| \left| \chi \left( p \right)\right| \ \leq \
  \phi_0 - \delta \right\} .
$$
Then
\begin{equation}
\int\limits_{M_\delta} dp\; p^\alpha \ \leq \ \frac{1}{\delta^2}
\int\limits_0^\infty 
dp\;p^\alpha \left| \chi \left( p \right) - \phi_0 \right|^2 \ < \ 
\frac{q_0^2 \varepsilon \left(q_0\right)}{\delta^2} \ .
\end{equation}

\eject

By property (B), 
\begin{eqnarray}
\infty &>& \int\limits_0^\infty dp \int\limits_0^\infty dq \frac{ q
  p^\alpha}{\sqrt{p^2+q^2}} \left| \phi \left( q,p\right)\right|^2 
\nonumber \\
&\geq& \int\limits_0^\infty dp \int\limits_{q_{0/2}}^{q_0} dq
\frac{\frac{q_0}{2} p^\alpha}{p+q_0} \left| \phi \left(
    q,p\right)\right|^2 \nonumber \\
&\geq& \int\limits_{M_\delta^c} dp \int\limits_{q_{0/2}}^{q_0} dq
\frac{\frac{q_0}{2} p^\alpha}{p +q_0} \left| \phi \left(
    q,p\right)\right|^2 \nonumber \\
&\geq& \frac{q_0^2}{4} \left( \phi_0 - \delta \right)^2
\int\limits_{M_\delta^c} dp \frac{p^\alpha}{p+q_0} \ . \nonumber
\end{eqnarray}
It follows that
$$
\int\limits_{M_\delta} dp \frac{p^\alpha}{p+q_0}\;+\;
\int\limits_{M_\delta^c} dp \frac{p^\alpha}{p+q_0}\;\leq\;
\frac{1}{q_0} \int\limits_{M_\delta} dp\,p^\alpha\;+\;
\int\limits_{M_\delta^c} dp \frac{p^\alpha}{p+q_0}\;<\;\infty \ . 
$$
This is a contradiction, since \ $M_\delta \cup M_\delta^c =
[0,\infty)$, and \ $\int dp \frac{p^\alpha}{p+q_0}$ diverges. 

Next, we consider the case $\alpha \leq - 1$. We change variables, $k
:= \frac 1 p$,
$$
dp\;=\;-\,\frac{1}{k^2} \;dk, \ \frac{\partial}{\partial p}\;=\;
-\, k^2 \;\frac{\partial}{\partial k} \ .
$$
Then conditions (A) -- (C) take the form
\begin{eqnarray}
&&(A')\quad \displaystyle\mathop{\rm lim}_{q\to 0 \atop q\in\Omega} \
\phi \left( q,k\right) \ = \ \phi_0\;>\;0\nonumber\\
&&(B')\quad \int\limits_0^\infty dk \int\limits_0^\infty dq \
\frac{q\;k^{\gamma-2}}{\sqrt{\left( \frac 1 k \right)^2 + q^2}} \
\left| \phi \left( q,k\right)\right|^2\;<\;\infty \nonumber\\
&&(C')\quad \int\limits_0^\infty dk \int\limits_0^\infty dq \
\frac{k^{\gamma-2}}{q} \ \left( k^4 \left| \phi,_k
    \left(q,k\right)\right|^2\;+\; \left| \phi,_q \left(
    q,k\right)\right|^2\right)\;<\;\infty ,\nonumber
\end{eqnarray}
where $\gamma := - \alpha > 0$.

Repeating the same arguments as above, we get again
\begin{equation}
\left( \frac{1}{q_0} \left| \chi \left( k\right) -
    \phi_0\right|\right)^2 \ \leq \int\limits_0^\infty \frac{dq}{q}
\left| \phi,_q \left( q,k\right)\right|^2 ,
\end{equation}
where $\chi (k)$ is the value of $\phi(q,k)$ at the minimum of $|\phi
(q,k)|$, for $q\in [0,q_0]$. By (C'), 
$$
\int\limits_0^\infty dk \int\limits_0^{q_0} dq \
\frac{k^{\gamma-2}}{q} \ \left| \phi,_q \left( q,k\right)\right|^2 \ <
\ \varepsilon' \left( q_0\right) \;<\;\infty ,
$$
with $\varepsilon'(q_0)\to 0$, as $q_0 \to 0$. Hence 
\begin{equation}
\int\limits_0^\infty dk\;  k^{\gamma-2} \left| \chi (k) - \phi_0\right|^2
\ \leq \ q_0^2 \varepsilon' \left( q_0\right) .
\end{equation}
Let $L_\delta \subseteq [ 0,\infty)$ be the set defined by
$$
L_\delta\;:=\; \left\{ k \biggm| \left| \chi (k)\right| \ \leq \ \phi_0
  - \delta\right\} .
$$
Then we have that
\begin{equation}
\int\limits_{L_\delta} dk\;k^{\gamma-2}\;\leq\; \frac{1}{\delta^2}
\int\limits_0^\infty dk \; k^{\gamma-2} \left| \chi(k)-\phi_0\right|^2
\;\leq\; \frac{q_0^2 \varepsilon' (q_0)}{\delta^2} \ .
\end{equation}
Condition (B') implies that 
\begin{eqnarray}
\infty &>& \int\limits_0^\infty dk\; k^{\gamma-2} \int\limits_0^\infty
\frac{q}{\frac 1 k + q} \left| \phi \left( q,k\right)\right|^2
\nonumber\\
&\geq& \int\limits_0^\infty dk\; k^{\gamma-2}
\int\limits_{q_{0/2}}^{q_0} \frac{q_{0/2}}{\frac 1 k + q_0} \left|
  \phi \left( q,k\right)\right|^2 \nonumber \\
&\geq& \int\limits_{L_\delta^c} dk\; k^{\gamma-1}
\int\limits_{q_{0/2}}^{q_0} \frac{q_{0/2}}{1+k\,q_0} \left( \phi_0 -
  \delta\right)^2 \nonumber \\
&\geq& \left( \frac{q_0}{2}\right)^2 \left( \phi_0 - \delta\right)^2
\int\limits_{L_\delta^c} dk \ \frac{k^{\gamma-1}}{1+k\;q_0} \ . 
\end{eqnarray}
Combining (3.47) and (3.48) we find that 
\begin{eqnarray}
&&\int\limits_{L_\delta} dk \
\frac{k^{\gamma-1}}{1+k\;q_0}\;+\;\int\limits_{L_\delta^c} dk \;
\frac{k^{\gamma-1}}{1+k\;q_0} \nonumber \\
&&\leq\;\frac{1}{q_0} \int\limits_{L_\delta} dk\; k^{\gamma-2}\;+\;
\int\limits_{L_\delta^c} dk \ \frac{k^{\gamma-1}}{1+k\;q_0}\;<\;\infty
\ .\nonumber 
\end{eqnarray}
This is a contradiction, because \ $L_\delta \cup L_\delta^c = [
0,\infty)$ \ and \ $\int dk\; \frac{k^{\gamma-1}}{1+k\;q_0}$ \
diverges for $\gamma \geq 1$.

This completes the proof  that functions satisfying
properties (A), (B) and (C) do not exist.

We have thus proven that eq.(3.1) only has the trivial solution \ $\Psi
= 0$. 

\vspace{1cm}

\section*{Acknowledgment}

\noindent We would like to thank H. Kalf for useful discussions.

\vspace{1cm}

\section*{References}

\begin{description}
\item[[1]] B. de Wit, J. Hoppe, H. Nicolai; Nuclear Physics B~\ub{305}
  [FS23] (1988) 545.
\item[[2]] J. Goldstone; unpublished. \\
J. Hoppe; MIT Ph.D. Thesis, 1982, \\
``Quantum Theory of a Relativistic
Surface'', in: ``Constraint's Theory and Relativistic Dynamics''
(eds. G. Longhi, L. Lusanna) Arcetri, Florence 1986, World
Scientific. 
\item[[3]] B. de Wit, M. L\"uscher, H. Nicolai; Nuclear Physics
  B~\ub{320} (1989) 135. 
\item[[4]] M. Baake, P. Reinicke, V. Rittenberg; Journal of
  Mathematical Physics~\ub{26} (1985) 1070. \\ R. Flume; Annals of
  Physics~\ub{164} (1985) 189.
\item[[5]] M. L\"uscher; Nuclear Physics~B~\ub{219} (1983) 233. \\
B. Simon; Annals of Physics~\ub{146} (1983) 209.
\item[[6]] T. Banks, W. Fischler, S.H. Shenker, L. Susskind; ``$M$
  Theory as a Matrix Model: a Conjecture'', hep-th/9610043.
\item[[7]] J. Hoppe; ``On Zero-Mass Bound-States in Super-Membrane
  Models'', hep-th/9609232.
\end{description}

\end{document}